\documentclass[aps,prd,showpacs,superscriptaddress,twocolumn,nofootinbib]{revtex4-2}

\usepackage{graphicx}
\usepackage{amsmath}
\usepackage{amssymb}
\usepackage{ulem}
\usepackage{bm}
\usepackage{float}
\usepackage{tikz}
\usepackage[colorlinks,
linkcolor=blue,
anchorcolor=blue,
urlcolor=blue,
citecolor=blue,
]{hyperref}

\begin{document}
\title{Photon Production by Chiral Magnetic Effect in the Early Stage of High-Energy Nuclear Collisions}
\author{Moran Jia}
\email{jiamr@ccnu.edu.cn}
\affiliation{Institute of Particle Physics and Key Laboratory of Quark and Lepton Physics (MOS), Central China Normal University, Wuhan 430079, China}

\date{\today}

\begin{abstract}
We present an event-by-event study of photons produced by the chiral magnetic effect in the early stage of high-energy nuclear collisions. We model the early stage with an evolving, 2+1 dimensional, glasma, initialized via the McLerran-Venugopalan model. The photons are produced from the interplay of the chiral anomaly of quantum chromodynamics and the strong magnetic fields generated by the colliding nuclei. In particular, we focus on the spectrum and the elliptic flow of the produced photons. We find that the chiral magnetic effect has a minor impact on the photon spectrum compared to other production mechanisms as well as to experimental data. However, the photons produced by the present early stage mechanism can exhibit a significant elliptic flow.
\end{abstract}
\maketitle

\section{introduction}\label{sec1}
The high-energy nuclear collision experiments at the Relativistic Heavy Ion Collider (RHIC) as well as the Large Hadron Collider (LHC) have produced a lot of interesting data about the medium formed in the collisions. There are convincing evidences in these collisions indicating the formation of quark-gluon plasma (QGP) which may persist for about $5$-$15$ fm$/c$ depending on the center-of-mass energy of the collision. Therefore, these experiments provide a unique opportunity to explore exotic manifestation of quantum chromodynamics (QCD) under extreme conditions of very high temperature. It is known that QCD contains classical gauge field configurations characterized by topologically invariant charges, which play crucial roles in, for example, the local breaking of parity ($\mathcal{P}$) and parity times charge conjugation ($\mathcal{CP}$) symmetries~\cite{RevModPhys.70.323}, and in the physics of the strong $\theta$ angle~\cite{RevModPhys.82.557}. The gauge field configurations with non-zero topological charges can also lead to chirality imbalance, inducing anomalous chiral effects coupled with the magnetic field ${\bm B}$ such as the Chiral Magnetic Effect (CME)~\cite{Kharzeev:2007tn,Kharzeev:2007jp,PhysRevD.78.074033,Fukushima_2013} and the Chiral Separation Effect (CSE)~\cite{PhysRevD.70.074018,PhysRevD.72.045011}. 

In non-central heavy-ion collisions, magnetic field is generated with its direction perpendicular to the reaction plane. The magnetic field is significantly stronger than those produced by laboratory magnets and its maximum strength can reach up to $O(100 m_{\pi}^2)$ ($\sim 10^{15}\,{\mathrm{T}}$)~\cite{Kharzeev:2007jp,PhysRevLett.36.517,SKOKOV_2009,Voronyuk:2011jd,Deng:2012pc,BZDAK2012171,PhysRevC.88.024911}. Thus, studies have triggered interest in exploring QCD matter under strong magnetic fields~\cite{PhysRevC.83.017901,PhysRevD.96.014023,PhysRevD.100.014020,Bali:2011qj,Bali:2012zg,Bali:2013esa,Bali:2014kia,Ding:2020hxw,Ding:2020inp}. Moreover, in heavy-ion collisions, the nonzero axial charge density can naturally arise in the system as explained later~\cite{Kharzeev:2001ev,Kharzeev:2004ey,Kharzeev:2007tn,Kharzeev:2007jp,Fukushima:2010vw}. The heavy-ion collisions offer a crucial environment for studying anomalous chiral effects. Investigations have been carried out both phenomenologically and experimentally towards finding direct evidence of these effects through chiral transport phenomena, see Refs.~\cite{Huang:2015oca,Kharzeev:2015znc,Liu:2020ymh,Li:2020dwr} for reviews. These anomalous chiral effects may also influence the production of direct photons in heavy-ion collisions~\cite{Basar:2012bp,Fukushima_2012,Huang:2018hgk,Tuchin:2019jxd}. Direct photons serve as one of the most versatile probes for studying early stages in high-energy heavy-ion collisions~\cite{David:2019wpt,Monnai:2022hfs}. Many authors have extensively investigated various photon production mechanisms and observables, particularly the transverse momentum spectrum and the elliptic flow $v_2$ of photons~\cite{Chaudhuri:2011up,vanHees:2011vb,Chatterjee:2012dn,Shen:2013vja,Monnai:2014kqa,Paquet:2015lta,Kim:2016ylr,Dasgupta:2018pjm,Garcia-Montero:2019kjk,Bhattacharya:2015ada,Linnyk:2015rco,Linnyk:2015tha,Greif:2016jeb,Vovchenko:2016ijt,Iatrakis:2016ugz,Oliva:2017pri,Hauksson:2017udm,Berges:2017eom,Monnai:2019vup,Kasmaei:2019ofu,Churchill:2020uvk,Wang:2020dsr,Jia:2022awu,Paquet:2022wgu,Sun:2023pil,Sun:2023rhh, Wang:2023fst}. Photons produced through the CME may serve as an indirect evidence of the chiral anomaly and could potentially contribute to large $v_2$ of direct photons observed in heavy-ion collisions.

In this article, we report on our study of the photon produced by the CME in the early stage of high-energy heavy-ion collisions, which can be described in terms of classical gluon fields, i.e., the evolving glasma, and the electromagnetic field produced by the colliding nuclei. The dynamics of the central rapidity region in the collision is dominated by gluons at small Bjorken $x$, at which the gluon saturation is expected to take place~\cite{Mueller:1999wm,PhysRevD.49.2233,PhysRevD.49.3352,PhysRevD.50.2225}.  The renormalization-group picture leads to a classical description of the Effective Field Theory of the Color Glass Condensate (CGC)~\cite{PhysRevD.55.5414,PhysRevD.54.5463,PhysRevD.59.014015,Iancu_20011,Iancu_20012}. The Lorentz-contracted colliding nuclei move almost along the light cone, with the large-$x$ partons acting as static sources of the small-$x$ modes that constitute the CGC fields, see Refs.~\cite{PhysRevD.49.2233,PhysRevD.49.3352,PhysRevD.50.2225,doi:10.1146/annurev.nucl.010909.083629,GELISCOLOR} for reviews. Right after the collision, color fields are fixed from the condition to avoid unphysical singularities. In such a way, longitudinal color-electric, $E_\eta^a$, and color-magnetic, $B_\eta^a$ ($a=1,2,\dots, N_c^2-1$), fields are formed. 
This initial transient state between the CGC and the quark-gluon plasma is called the glasma~\cite{Lappi:2006fp} which evolves in proper time according to the Classical Yang-Mills (CYM) equations~\cite{PhysRevC.89.024907,FUKUSHIMA2012108,PhysRevLett.111.232301,Berges:2020fwq,Gale:2012rq,Gale:2012in,PhysRevD.97.076004}. We use the McLerran-Venugopalan (MV) model to initialize our approach, where the evolving glasma enjoys a nontrivial feature, among other interesting properties, that is a locally nonzero $\mathrm{Tr} ({\bm E}\cdot {\bm B})$~\cite{Kharzeev:2005iz,Lappi:2006fp,Lappi:2017skr,Guerrero-Rodriguez:2019ids,jia2020fluctuations}. This inner product of the vector and the axial vector is parity odd, leading to a chiral imbalance, namely an imbalance in the number of left-handed and right-handed quarks via the chiral anomaly in QCD\@. The coupling of the chirally imbalanced medium to the magnetic field gives rise to the CME, as well as the photon production, as we discuss in this paper. Hence, photons can be produced by the CME in the early stage of high-energy heavy-ion collisions. The main novelty of our study is that we present an approach of the production rate of photons via CME, for which we can carry out event-by-event analysis from the glasma calculation, in particular the color-$\mathrm{Tr} ({\bm E}\cdot {\bm B})$, to locally compute the chiral density. We then combine this with a profile for the electromagnetic field, and use the CME to extract the photon spectrum and photon elliptic flow~\cite{Fukushima_2012}.

The article is organized as follows. In Sec.~\ref{sec2}, we briefly review the CME in an evolving glasma. In Sec.~\ref{sec3}, we introduce chiral-anomaly photon production in the presence of a strong magnetic field in the initial stage. In Sec.~\ref{sec4}, we show our numerical results for both the spectrum and the elliptic flow of the photon. Finally, in Sec.~\ref{sec5}, we summarize our results. Throughout this article, we rescale the gauge fields by the QCD coupling, namely $A_{\mu}\rightarrow A_{\mu}/g$, thus $g$ does not appear explicitly in the calculations.

\section{Chirality production in the evolving glasma}\label{sec2}

The glasma initial condition accommodates parallel ${\bm E}$ and ${\bm B}$ which can be a source for the chirality production through the chiral anomaly.  We will review basic formulas for the glasma evolution and show the resultant chiral chemical potential as a function of the proper time.

\subsection{The evolving glasma}\label{sec2.1}

In the MV model, the color charge densities $\rho^a$ act as  static sources of the transverse CGC fields in the two colliding nuclei. They are assumed to be random variables that, for each nucleus, are normally distributed with zero average and with variance specified by
\begin{equation}
    \langle\rho^a(\bm{x}_{T},\eta_1)\rho^b(\bm{y}_{T},\eta_2)\rangle=
	(g^2\mu)^2\delta^{ab}\mathbf{\delta}(\bm{x}_{T}-\bm{y}_{T})
	\delta(\eta_1-\eta_2),\label{eq2.2.1}
\end{equation}
where $a$ and $b$ are the adjoint color indices, $\bm{x}_{T}$ and $\bm{y}_{T}$ denote transverse plane coordinates, and $\eta_1$, $\eta_2$ are the spacetime rapidities. In the MV model, $g^2\mu$ is the only energy scale which is related to the saturation momentum $Q_s$. We refer to the estimate of~Ref.~\cite{LappiWilson}, namely, $Q_s/g^2\mu\approx 0.57$. Due to the $\delta$-functions in Eq.~\eqref{eq2.2.1} the color charge densities are uncorrelated in transverse plane as well as in rapidity. To specify the glasma initial condition,
it is convenient to work in Milne coordinates $(\tau,\eta)$, where
\begin{align}
	t &= \tau \rm cosh\eta,\label{eq2.2.2}\\
	z &= \tau \rm sinh\eta,\label{eq2.2.3}
\end{align}
and in the radial gauge, $A_{\tau}=0$. In order to compute the glasma fields we firstly solve the Poisson equations, namely
\begin{align}
	-{\bm\nabla}^2 \alpha^{+(A)}(x^-,\bm{x}_{T}) &=\rho^{(A)}(x^-,\bm{x}_{T}),\label{eq2.2.4}\\
	-{\bm\nabla}^2 \alpha^{-(B)}(x^+,\bm{x}_{T}) &=\rho^{(B)}(x^+,\bm{x}_{T}),\label{eq2.2.5}
\end{align}
with $A$ and $B$ denoting the two colliding nuclei. The solutions of these equations are simply
$\alpha^{\pm (A/B)}=-{\bm\nabla}^{-2}\rho^{(A/B)}$.  Then, they are gauge rotated as
\begin{align}
	\alpha^{(A)}_{i}(\bm{x}_{T}) &=iU^{(A)}(\bm{x}_{T})\partial_iU^{(A)\dagger}(\bm{x}_{T}),\label{eq2.2.6}\\
	\alpha^{(B)}_{i}(\bm{x}_{T}) &=iU^{(B)}(\bm{x}_{T})\partial_iU^{(B)\dagger}(\bm{x}_{T}),\label{eq2.2.7}
\end{align}
where the Wilson lines are defined as
\begin{align}
    U^{(A)\dag}(\bm{x}_{T}) &\equiv \mathcal{P}\exp\left[-i\int dz^- \alpha^{+(A)}(z^-,\bm{x}_{T})\right],\\
    U^{(B)\dag}(\bm{x}_{T}) &\equiv \mathcal{P}\exp\left[-i\int dz^+ \alpha^{-(B)}(z^+,\bm{x}_{T})\right]
\end{align}
with $\mathcal{P}$ being the path order operator and $z(\bm{x}_{T})$ a trajectory. In terms of these fields, the glasma gauge potential at $\tau\rightarrow 0^+$ can be written as~\cite{PhysRevD.52.6231,PhysRevD.52.3809}:
\begin{eqnarray}
	&A_{i}=\alpha_{i}^{(A)}+\alpha_{i}^{(B)}, i=x,y,\label{eq2.2.8}\\
	&A_{\eta}=0.\label{eq2.2.9}
\end{eqnarray}

Solving the Yang-Mills equations near the light cone, one finds that the transverse color electric and color magnetic fields vanish as $\tau\rightarrow 0$, but the longitudinal electric and magnetic fields are non-vanishing \cite{Fries_2006}:
\begin{eqnarray}
	&E_{\eta}=i\sum_{i}[\alpha_{i}^{(A)},\alpha_{i}^{(B)}], \label{eq2.2.10}\\
	&B_{\eta}=i([\alpha_{x}^{(A)},\alpha_{y}^{(B)}]+[\alpha_{x}^{(B)},\alpha_{y}^{(A)}]).\label{eq2.2.11}
\end{eqnarray}
In all the discussion above, we have neglected the possibility of fluctuations that, among other things, would break the longitudinal boost invariance. While these are relevant for the onset of the hydrodynamical flow~\cite{Dusling:2010rm,Romatschke:2006nk,Romatschke:2005ag,Romatschke:2005pm,PhysRevLett.111.232301,PhysRevD.97.076004,FUKUSHIMA2012108,PhysRevC.89.024907}, they are not crucial for the production of chiral density and for the evolution of the topological charge density. We also assume that $g^2\mu$ is the same in the two colliding nuclei and it has no dependence on the transverse plane coordinates. We will remove these assumptions in the future, in order to mimic the energy density profile that would be produced in realistic collisions~\cite{Gale:2012rq,Gale:2012in} and to incorporate a full 3+1 dimensional glasma simulation~\cite{Schlichting:2020wrv,Matsuda:2023gle,Matsuda:2024moa,Ipp:2024ykh,McDonald:2023qwc,Ipp:2021lwz,Ipp:2020igo,Avramescu:2023qvv}.

After preparing the initial condition of CYM equations, within the gauge $A_{\tau}=0$ the Lagrangian density reads
\begin{equation}
	\mathcal{L}=\frac{1}{2}{\rm Tr}\big[-\frac{2}{\tau^2}(\partial_{\tau}A_{\eta})^2-2(\partial_{
		\tau}A_{i})^{2}+\frac{2}{\tau^2}F^{2}_{\eta i}+F_{ij}^{2}\big],\label{eq2.2.12}
\end{equation}
and the canonical momenta are given by
\begin{eqnarray}
	E_{i}=\tau\partial_{\tau}A_{i},\label{eq2.2.13}\\
	E_{\eta}=\frac{1}{\tau}\partial_{\tau}A_{\eta}.\label{eq2.2.14}
\end{eqnarray}
As a consequence, the Hamiltonian density is
\begin{equation}
	\mathcal{H}={\rm Tr}\big[\frac{1}{\tau^2}E_{i}^2+E_{\eta}^2+\frac{1}{\tau^2}F_{\eta i}^2+\frac{1}{2}F_{ij}^2\big].\label{eq2.2.15}
\end{equation}
The CYM equations in Milne coordinates are
\begin{eqnarray}
	&\partial_{\tau}E_{i}=\frac{1}{\tau}\mathcal{D}_{\eta}F_{\eta i}+\tau\mathcal{D}_{j}F_{ji},\label{eq2.2.16}\\
	&\partial_{\tau}E_{\eta}=\frac{1}{\tau}\mathcal{D}_{j}F_{j\eta},\label{eq2.2.17}
\end{eqnarray}
where $\mathcal{D}_{\mu}=\partial_{\mu}+iA_{\mu}$ is the covariant derivative. We define $B_i=-\epsilon^{ij}F_{j\eta}$ and $B_{\eta}=-\frac{1}{2}\epsilon^{ij}F_{ij}$
as, respectively, the $i$ and $\eta$ components of the color magnetic field.

\subsection{The chiral anomaly}\label{sec2.2}

As mentioned in Sec.~\ref{sec1}, immediately after the collision, nonzero longitudinal components of the gauge fields as well as the quark masses $m_q$ lead to the non-conservation of the U(1) axial symmetry of QCD, according to the Adler-Bell-Jackiw anomaly equation~\cite{PhysRev.177.2426,Bell:1969ts}:
\begin{equation}
\partial_{\mu}j^{\mu}_{5} = 2m_q\bar{q}i\gamma_{5}q - \frac{1}{16\pi^2}F_{\mu\nu}^{a}\tilde{F}_{a}^{\mu\nu}\label{eq2.1.1},
\end{equation}
where $F_{\mu\nu}^{a}=\partial_{\mu}A_{\nu}^{a}-\partial_{\nu}A_{\mu}^{a} + f_{abc}A_{\mu}^{b}A_{\nu}^{c}$ denotes the non-Abelian field strength tensor and $\tilde{F}_{a}^{\mu\nu} = 1/2\epsilon^{\mu\nu\rho\sigma}F_{\rho\sigma}^{a}$ is its dual~\cite{PhysRevLett.37.8}. In the massless limit, the anomaly equation for each flavor takes form as:
\begin{equation}
\partial_{\mu}j^{\mu}_{5} = -\frac{1}{16\pi^2}F_{\mu\nu}^{a}\tilde{F}_{a}^{\mu\nu}\label{eq2.1.2}.
\end{equation}
It is convenient to use the definitions as in \cite{jia2020fluctuations} and write Eq.~\eqref{eq2.1.2} in Milne coordinates:
\begin{equation}
\left(\partial_\tau + \frac{1}{\tau}\right)j_5^\tau
=  \frac{1}{8\pi^2}{\rm Tr}\left(\bm E\cdot \bm B\right)\label{eq2.1.3}.
\end{equation}
In writing Eq.~\eqref{eq2.1.3} we have used the 
boost-invariance approximation as well as 
the simplifying assumption 
that the divergence of the axial current in the transverse plane vanishes; the latter implies that there is no net flux of the axial current across the transverse plane, in agreement with the common lore of neglecting the transverse expansion in the early stage.
Here, the color-electric field $\bm E = E_a T_a$ and the color-magnetic field $\bm B = B_a T_a$ are those of the evolving glasma (here and in next subsection we use boldface to denote vectors in color space). The color generators are normalized as 
${\rm Tr}(T_{a}T_{b})=2\delta_{ab}$. 

\subsection{The chiral chemical potential}\label{sec2.3}

In this paper we define $n_5:=j_{5}^{\tau}$, which is the chiral density in a 3-dimensional volume of a cell with extension $d^2x_\perp$ in the transverse plane and $\tau d\eta$ in the longitudinal direction. From Eq.~\eqref{eq2.1.3} we obtain:
\begin{equation}
	 n_5(\tau,x_\perp)=\frac{1}{8\pi^2}\frac{1}{\tau}\int_0^\tau d\tau^\prime \tau^\prime~ {\rm Tr}\left(
	 \bm E \cdot \bm B
	 \right) .\label{eq2.3.1}
\end{equation}
A natural way to introduce chirality in effective models is to include a chiral chemical potential $\mu_{5}$ conjugated to chiral density~\cite{PhysRevD.81.114031,PhysRevD.83.105008,PhysRevD.84.014011,Gatto:2011wc}. By performing a U(1) rotation, we can identify $\theta F_{a}^{\mu\nu}\tilde{F}^{a}_{\mu\nu}$ with a fermionic-like contribution $\frac{1}{2N_f}\partial_{\mu}\theta\bar{\psi}\gamma^{\mu}\gamma^{5}\psi$. Thus, when compared to the operator $\mu_{5}\bar{\psi}\gamma^0\gamma^5\psi$, the chiral chemical potential can be defined as $\mu_{5}=\partial_0\theta/2N_f$ \cite{PhysRevD.78.074033}, where $\theta$ works as a spacetime variable.

Once $n_5$ is known, we need to compute $\mu_5$. To simplify this computation, we assume that the relation between the two follows that of a massless fermion gas at zero temperature:
\begin{equation}
	n_5=  N_f N_c\frac{\mu_{5}^3}{3\pi^2}.\label{eq3.11}
\end{equation}

\begin{figure}[h!]
	\centering
	\includegraphics[width=\linewidth]{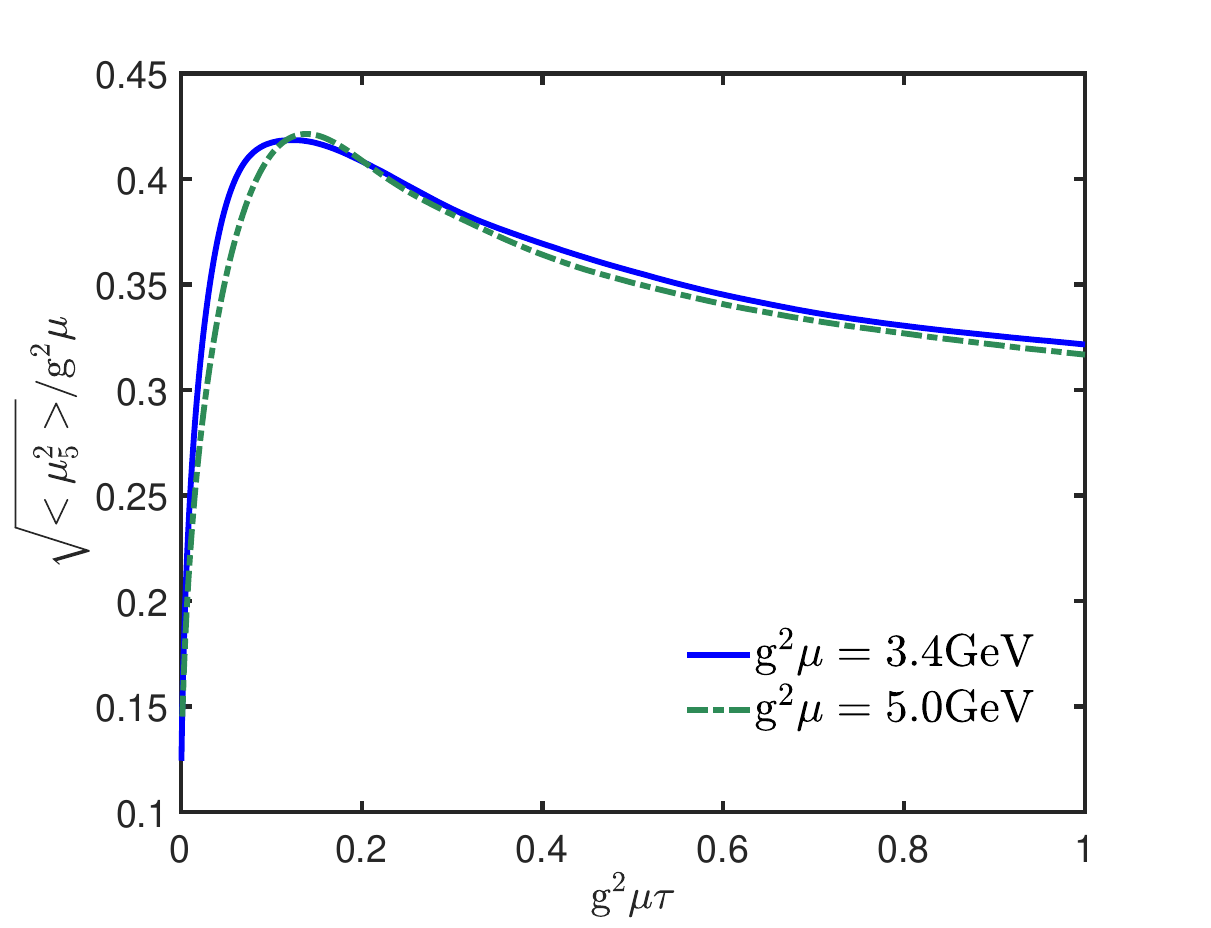}
\caption{
 $\sqrt{\left\langle\mu_{5}^2\right\rangle}/g^2\mu$ as a function of $g^2\mu\tau $, for two values of $g^2\mu$. The lattice spacing is setting as a =0.02 $\rm fm$ and the results are averaged over 100 events.
}
\label{cmemu5}
\end{figure}

In Fig.~\ref{cmemu5}, we plot $\sqrt{\langle \mu_5^2\rangle}$ in units of $g^2\mu$, versus $g^2\mu\tau$ for two representative values of $g^2\mu$, as obtained within our calculations. Here, $\left\langle\dots\right\rangle$ denotes ensemble average as well as average over transverse plane. It is interesting to notice that $\sqrt{\left\langle\mu_{5}^2\right\rangle}$ is of the order of $g^2\mu$ as expected, since $g^2\mu$ is the only energy scale in the model. Furthermore, the values of $\sqrt{\left\langle\mu_{5}^2\right\rangle}$ that we find within our glasma calculations are in the same ballpark of the estimates of~\cite{Frenklakh:2024jff}.

\section{The photons produced by chiral magnetic effect}\label{sec3}

The neutral pseudoscalar field $\theta(x)$ can couple to photons, leading to a single photon emission, namely $\theta + \gamma \rightarrow \gamma$~\cite{Basar:2012bp,Fukushima_2012}, where, within our context, the photon on the left-hand side is provided by the time-dependent background magnetic fields produced in high energy nuclear collisions~\cite{Fukushima_2012,PhysRevC.67.025203}. In this work, $\theta(x)$ is related to the evolving glasma fields as discussed in the previous section. The effective coupling between the photon and the $\theta$ is given by~\cite{Fukushima_2012}
\begin{equation}
\mathcal{L}_P=\frac{N_ce^2\rm {\rm Tr}(Q^2)}{8N_f\pi^2}\epsilon^{\mu\nu\rho\sigma}[\mathcal{A}_{\mu}\bar{F}_{\nu\rho}+\mathcal{A}_{\mu}(\partial_{\nu}\mathcal{A}_{\rho})]\partial_{\sigma}\theta,
\label{eq:analeq0}
\end{equation}
where $\mathcal{A}^\mu$ denotes the photon field, $\bar F^{\mu\nu}$ is the strength tensor of the background electromagnetic field, and $Q=\mathrm{diag}\{2/3,-1/3,-1/3\}$ is the electric-charge matrix in quark flavor space. The second term in the right hand side of Eq.~\eqref{eq:analeq0} represents the two-photon production process, $\theta\rightarrow\gamma\gamma$, while the first one corresponds to the reverse Primakoff effect, $\theta+B\rightarrow\gamma$, involving the background field strength $\bar F_{\mu\nu}$. Here we are interested in the situation where the background field is so strong that we can neglect the contribution from the second term.

Using $\mu_5=\partial_0\theta/2N_f$ and defining
\begin{equation}
\begin{aligned}
\zeta_i(\bm q)\equiv\int d^4x\, e^{-iq\cdot x}eB_i(x)\mu_5(x)\,,~~i=x,y\,,\label{eq1.2}
\end{aligned}
\end{equation}
the unpolarized spectrum of photons produced by magnetic field $B_{i}$ and $\theta$ is given by
\begin{equation}
\begin{aligned}
q_0\frac{dN_{\gamma}}{d^3q}=&q_0\sum_{i}|\mathcal{M}(i;\bm q)|^2\\
=&\alpha_e\left(\frac{N_c{\rm Tr}(Q^2)}{4\pi^3}\right)^2\bigg|(1-\frac{q_x^2}{\bm q^2})\zeta_x^2(\bm q)\\
&+(1-\frac{q_y^2}{\bm q^2})\zeta_y^2(\bm q)-\frac{2q_xq_y}{\bm q^2}\zeta_x(\bm q)\zeta_y(\bm q)\bigg|\,.\label{eq1.1}
\end{aligned}
\end{equation}
Here $\alpha_e\equiv 1/4\pi$, while the electromagnetic coupling $e$ is absorbed in the definition of $\zeta_i(\bm q)$. In Milne coordinates, Eq.~\eqref{eq1.2} reads
\begin{equation}
\begin{aligned}
\zeta_i(\bm q)=\int d\tau d\eta\, &\tau 
e^{-i(q_{T}\tau\cosh(y-\eta))}
\int d^2x_{T}\,e^{i\bm q_{T}\cdot \bm x_{T}}\\
&\times eB_i(\tau,{\bm x}_{T},\eta)\mu_5(\tau,{\bm x}_{T}),\label{eq1.5}
\end{aligned}
\end{equation}
where $y$ is the momentum rapidity. Finally, we notice that $\mu_5$ is assumed to be $\eta$-independent due to the boost invariance of the evolving glasma fields.
 
\section{Results}\label{sec4}

In this section, we present our event-by-event results with all parameters keep the same as in Fig.~\ref{cmemu5}. Firstly, we explain the simplified magnetic field profile utilized in our calculations. Then, we study the photon spectrum produced by the CME in different scenarios and make a comparison with other sources of direct photons, namely prompt and early-stage photons. Finally, we comment on the elliptic flow of the CME photons.

\subsection{The magnetic field profile}\label{sec4.1}

Since for current calculations we confine our focus to the overlapping region of the two colliding nuclei, a fairly good approximation for the magnetic field is that of a homogeneous field~\cite{Voronyuk:2011jd,Deng:2012pc}, whose dominant component is along the $y$-axis. Additionally, we ignore the electric field as it does not contribute to the production of CME photons.

While some numerical simulations have demonstrated that the decay of electromagnetic fields during the early stage is influenced by the medium formed in collisions~\cite{Yan:2021zjc,Wang:2021oqq,Zhang:2022lje}, our understanding of the complete evolution of electromagnetic fields in heavy-ion collisions remains incomplete. Therefore, in order to mimic the time evolution in the early stage, we implement the magnetic field as $B_y=B_0f(\tau)\cosh{\eta}$, where we introduce the time-dependent functions~\cite{Sun:2020wkg}
\begin{eqnarray}
&&{\rm Case}~1:~f(\tau) = \left(1+\tau^2/\tau_B^2\right)^{-1}\,,\label{eq4.1.2}\\
&&{\rm Case}~2:~f(\tau) = \left(1+\tau/\tau_B\right)^{-1}\,,\label{eq4.1.3}
\end{eqnarray}
with $\tau_B$ being the lifetime of the decaying magnetic field. The estimation in Ref.~\cite{Sun:2020wkg} showed that $\tau_B=0.4\pm0.1$ fm/c for the case 2 by examining the directed flow splitting of neutral $D$ mesons.
In the present work we use a couple of representative values of $\tau_B$ in the aforementioned range. 
$B_0$ is the averaged value of $|\left\langle B_y\right\rangle|$ on the transverse plane at $\tau=0^+$. 
We choose $eB_0=50\,m_{\pi}^2$,
that agrees with the estimate for
Pb-Pb collisions at $\sqrt{s_{NN}}=2.76$ GeV with impact parameter $b=8$ fm~\cite{Deng:2012pc,Oliva:2020mfr}, and later we will use it in approximations of 20\%\textminus 40\% central collisions.

With the simplification on the EM fields explained above, Eq.~\eqref{eq1.1} becomes:
\begin{equation}
q_0\frac{dN_{\gamma}}{d^3q}=\alpha_e\left(\frac{N_c{\rm Tr}(Q^2)}{4\pi^3}\right)^2(1-\frac{q_y^2}{q^2})\left\langle{|\zeta_y(\bm q_T)|^2}\right\rangle\,,\label{eq4.2.1}
\end{equation}
where $\left\langle\dots\right\rangle$ denotes the ensemble average. According to Eq.~\eqref{eq1.5} we have, within our approximations,
\begin{equation}
\begin{aligned}
&\left\langle{|\zeta_y(\bm q_T)|^2}\right\rangle=\int \tau_1 d\tau_1 d\eta_1 \int \tau_2 d\tau_2 d\eta_2 \\
&\times e^{-iq_{T}(\tau_1\cosh{\eta_1}-\tau_2\cosh{\eta_2})}eB_y(\tau_1,\eta_1)eB_y(\tau_2,\eta_2)\\
&\times\int d^2x_{T}d^2y_{T}\left\langle\mu_5(\tau_1,\bm{x}_{T})\mu_5(\tau_2,\bm{y}_{T})\right\rangle 
e^{i\bm q_{T}\cdot 
(\bm x_{T} - \bm y_{T})}.\label{eq4.2.2}
\end{aligned}
\end{equation}
Moreover, the $\mu_5$ - correlator in present evolving glasma fields is invariant under rotations in the transverse plane, which implies that $\left\langle{|\zeta_y(\bm q_T)|^2}\right\rangle$ above depends on $q_T=|\bm q_T|$ only. Noticing that $d^3q/q_0 = d^2\bm q_T dy$ with $y=(1/2){\rm ln}[(q_0+q_z)/(q_0-q_z)]$ the momentum rapidity, averaging over the azimuthal angle
at $y=0$
we finally get 
\begin{eqnarray}
\left.\frac{dN_\gamma}{2\pi q_T dq_T dy}\right|_{y=0}
&=&\frac{1}{2}
\alpha_e\left(\frac{N_c{\rm Tr}(Q^2)}{4\pi^3}\right)^2\left\langle{|\zeta_y(\bm q_T)|^2}\right\rangle.\nonumber\\
&&
\label{eq:analeq1}
\end{eqnarray}
For the sake of notation, from now on we will not 
rewrite the subscript $y=0$ in our equations: it is
understood that we consider only photons at 
midrapidity.

\subsection{The spectrum of the CME photons}\label{sec4.3}

In this subsection, we present the $q_T$-spectrum of the CME photons. We choose $g^2\mu=3.4$ GeV that approximately corresponds to the value relevant for modeling the Pb-Pb collision at $2.76$ TeV, but we checked that changing this parameter qualitatively leads to the same results.

\begin{figure}[t!]
	\centering
	\includegraphics[width=\linewidth]{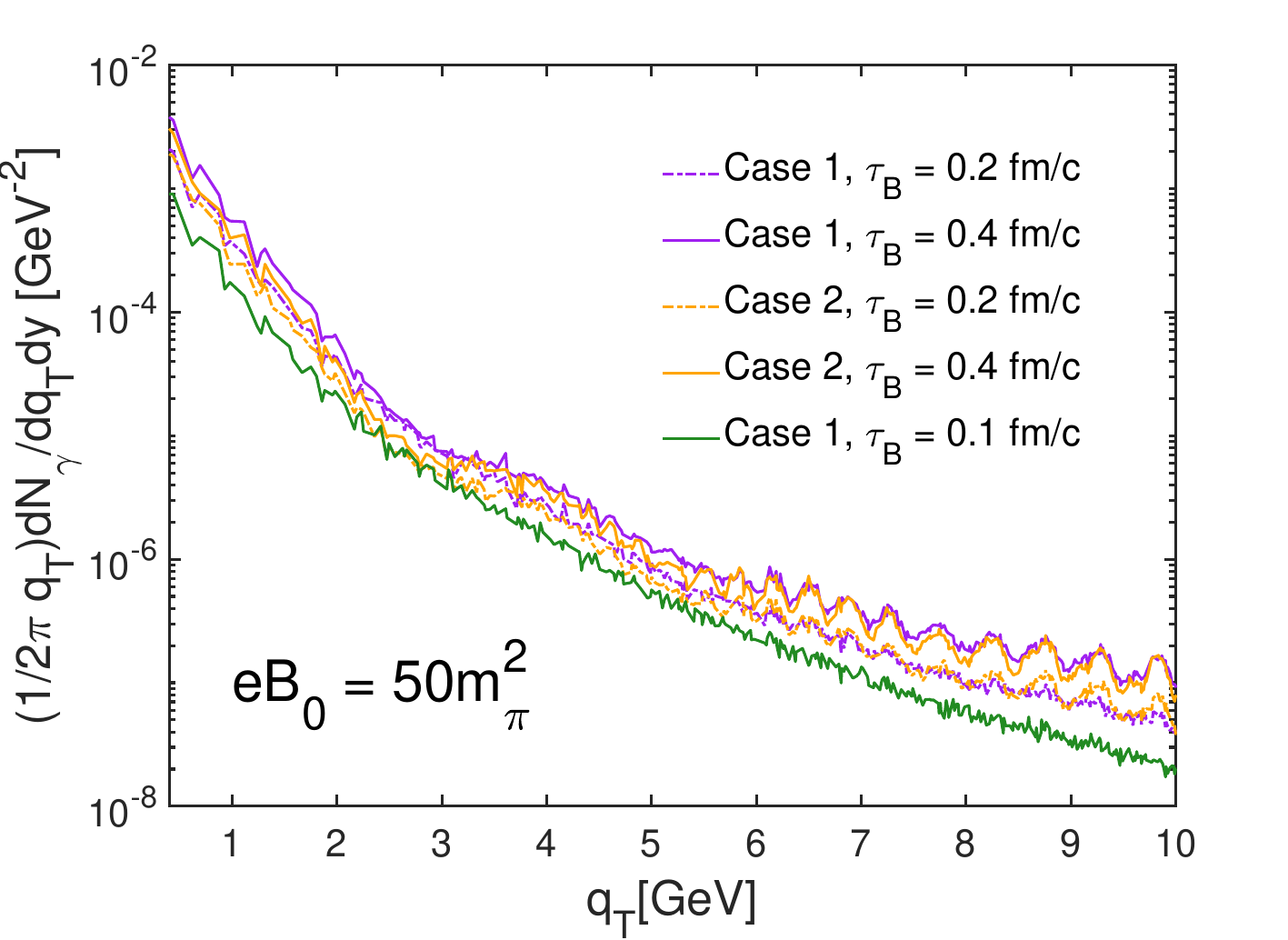}
	\caption{The spectrum of the CME photons 
 at midrapidity
 for different cases with different $\tau_B$, corresponding to Eq.~\eqref{eq:analeq1}.
 Results correspond to $\tau=0.5$ fm/c.}
	\label{Fig:photo2}
\end{figure}

In Fig.~\ref{Fig:photo2}, we plot the CME-photon spectrum obtained with the parametrizations of the magnetic field given by Eqs.~\eqref{eq4.1.2}-\eqref{eq4.1.3} with different $\tau_B$; particularly, we use $\tau_B=0.2$ fm/c and $\tau_B=0.4$ fm/c following the estimates of Ref.~\cite{Sun:2020wkg}. We used $eB_0=50 m_\pi^2$ as an estimate of the initial magnetic field~\cite{Deng:2012pc,Oliva:2020mfr} and later for approaches of 20\%\textminus40\% central collisions. The results have been obtained after averaging over $100$ events
at the proper time $\tau=0.5$ fm/c. 
We notice that the CME-photon spectrum is not a thermal one, rather it decays at large $q_T$ as a power-law; the specific power can be extracted within a simple ansatz on the correlations of $\mu_5$, as we discuss in Sec.~\ref{sec4.2}.

We also notice that changing from Case 1 to Case 2 results in a slight lowering of the photon multiplicity. This little difference can be understood by noticing that the magnetic field corresponding to
Case 1 decays slower than that in Case 2 in the small $\tau$ region, hence the CME is more efficient in producing photons in the former case rather than in the latter. However, the difference between the two cases is very small, as we can see in the figure where the data corresponding to the two cases are barely distinguishable from each other. Moreover, doubling $\tau_B$ leads to a difference of about 50\% at most. Given the small difference in the spectrum between Case 1 and Case 2 across various values of $\tau_B$, we will restrict ourselves to Case 1 with $\tau_B=0.4$ fm/c in the rest of the paper.

\begin{figure}[t!]
	\centering
	\includegraphics[width=\linewidth]{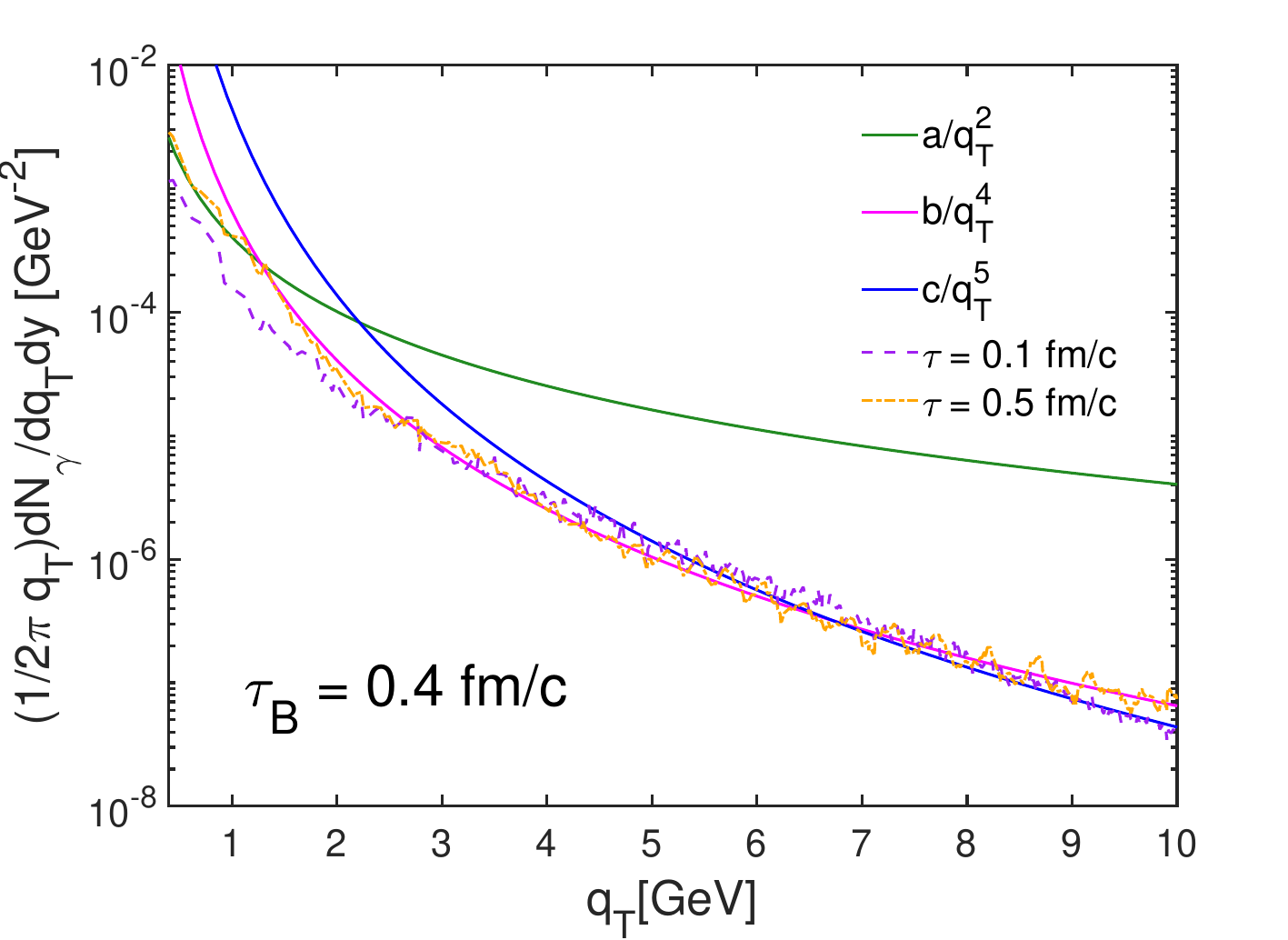}
	\caption{The spectrum of the CME-photons for $\tau_B=0.4$ fm/c, computed at $\tau=0.1$ fm/c (dashed purple) and $\tau=0.5$ fm/c (dot-dashed orange line). Representative power laws are also shown to guide the eye. }
	\label{Fig:photo3}
\end{figure}

In Fig.~\ref{Fig:photo3}, we show the CME-photon spectrum for $\tau_B=0.4$ fm/c, computed at $\tau=0.1$ fm/c (dashed purple line) and $\tau=0.5$ fm/c (dot-dashed orange line). Additionally, we include some representative power laws to guide the eye. The value of $\tau_B$ was chosen in order to illustrate the difference between the photon spectrum computed at early and late times. It is evident that the representative power laws of $a/q_T^4$ and $b/q_T^5$ deviate from the computed spectrum at small $q_T$, which we will analyze through an analytical approach in Sec.~\ref{sec4.2}.

\begin{figure}[t!]
	\centering
	\includegraphics[width=\linewidth]{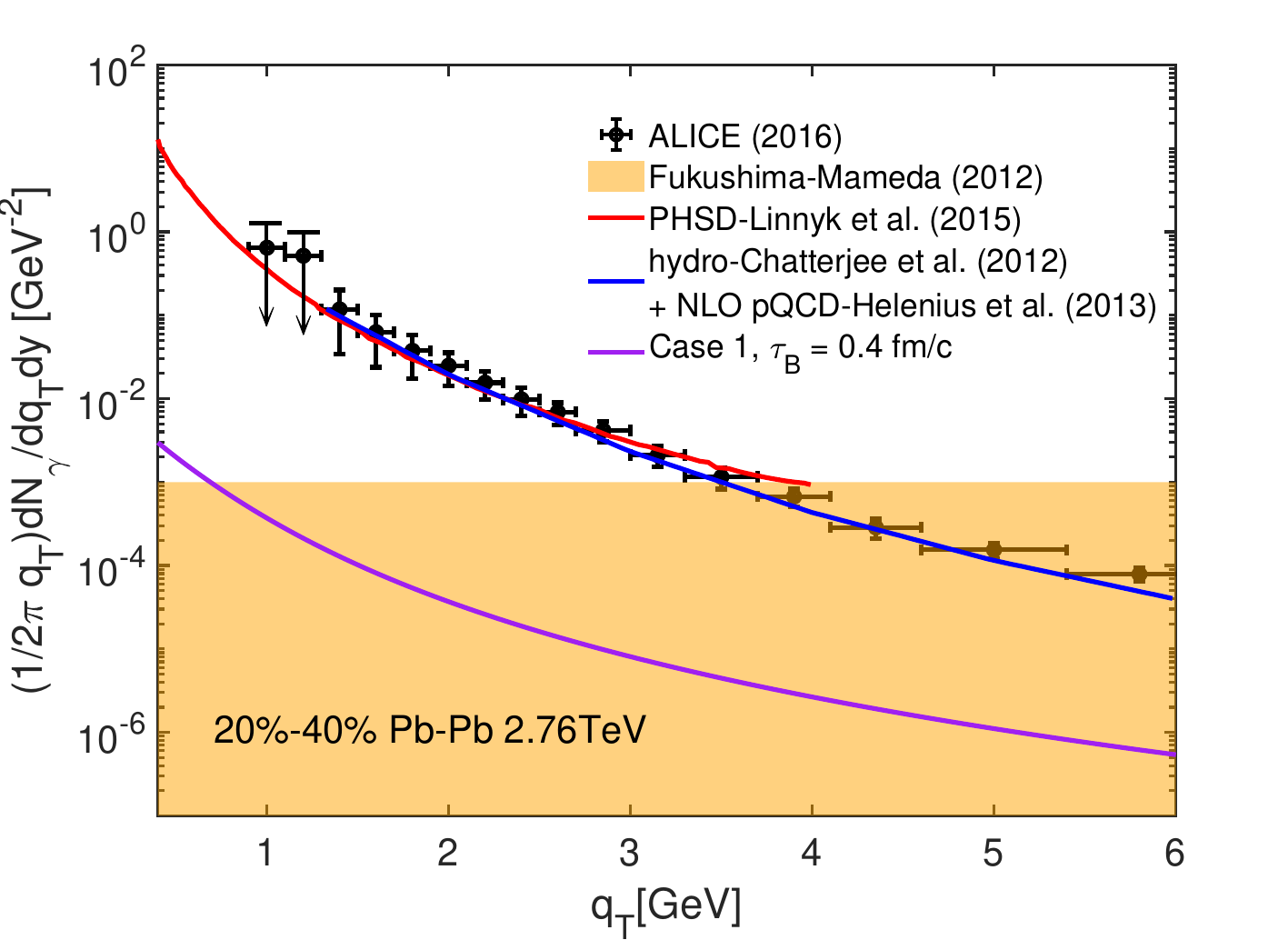}
	\caption{Comparison between CME photons, measurement by ALICE \cite{ALICE:2015xmh}, prompt photons with PHSD approach \cite{Linnyk:2015tha} and thermal photons from the hydrodynamic model \cite{Chatterjee:2012dn} with NLO pQCD computation \cite{Helenius:2013bya}.}
	\label{Fig:photo4}
\end{figure}

It is interesting to make a comparison of the spectrum of CME photons with that of photons produced by other sources in the early stage of the collision. In Fig.~\ref{Fig:photo4} we compare the fitted CME photon spectrum (solid purple line) with the spectrum of measured by ALICE \cite{ALICE:2015xmh}, prompt photons computed within the Parton-Hadron-String Dynamics (PHSD) approach~\cite{Linnyk:2015tha} (solid red line) and the photon spectrum obtained from hydrodynamic model with NLO pQCD computation~\cite{Chatterjee:2012dn, Helenius:2013bya} (solid blue line). The orange band corresponds to the estimate for the magnitude of the CME photons done in Ref.~\cite{Fukushima_2012}. Overall, the spectrum of the CME photons is lower than that of the prompt ones. Nevertheless, they still can contribute to the elliptic flow, as we discuss in Sec.~\ref{secv2}.

\subsection{The analysis with the $\mu_{5}$-correlator}\label{sec4.2}

Besides the full numerical calculation shown in the previous subsections, it is possible to capture some properties of the CME-photon spectrum through an analysis that makes use of the $\mu_5$-correlator discussed in Sec.~\ref{sec2.3}.
The analysis presented here is not intended to be quantitative; rather, its purpose 
is to extract the $q_T$-behavior of the CME-photon spectrum at small and large $q_T$, and relate it to the shape of the $\mu_5$-correlator
as well as to the duration of the magnetic field.

To begin with, we introduce an ansatz for the $\mu_5$-correlator needed in Eq.~\eqref{eq4.2.2}, namely,
\begin{equation}
\begin{aligned}
\left\langle\mu_5(\tau_1,\bm{x}_{T})\mu_5(\tau_2,\bm{y}_{T})\right\rangle&= \left\langle\mu_5^2
(\tau_1)
\right\rangle 
\mathcal{F}\left(\frac{u}{\lambda}\right)
\lambda_{\tau} \delta(\tau_1-\tau_2).\label{eq4.2.3}
\end{aligned}
\end{equation}
Here $\lambda_{\tau}$ stands for a correlation time scale, 
needed to balance the dimension carried by $\delta(\tau_1-\tau_2)$,
$u=|\bm x_T - \bm y_T|$ is the distance in the
transverse plane, and $\lambda$ denotes a decay length
of the correlations in the transverse plane, $\lambda=O(1/g^2\mu)$, see \cite{jia2020fluctuations}. In writing the ansatz~\eqref{eq4.2.3}, we have assumed that correlations in time decay very quickly, as we have
confirmed in full numerical simulations which show that for every practical purpose we can assume $\lambda_\tau\ll\lambda$, we checked that changing the $\delta$-function with an exponential decay in time does not qualitatively change the results discussed below. Besides, we require that $\mathcal{F}\left(\frac{u}{\lambda}\right)$ die quickly at large distance, which is inspired by \cite{jia2020fluctuations}.

Substituting Eq.~\eqref{eq4.2.3} into Eq.~\eqref{eq:analeq1} we easily get
\begin{eqnarray}
\frac{dN_{\gamma}}{2\pi q_{T}dq_{T}dy}&=&
\frac{1}{2}
\alpha_e\left(\frac{N_c{\rm Tr}(Q^2)}{4\pi^3}\right)^2
\nonumber\\
&&\times  \lambda_{\tau} (eB_0)^2A_{T}\mathcal{J}(q_{T},\tau).\label{eq:analeq1.1}
\end{eqnarray}
Here $\tau$ denotes the proper time at which the spectrum is computed, $A_T$ denotes the transverse area, $A_{T}=\int d^2(\frac{\bm x_{T}+\bm y_{T}}{2})$,
and $\bm u= \bm{x}_{T}-\bm{y}_{T} $. Function $\mathcal{J}(q_{T})$ is defined as
\begin{equation}
\begin{aligned}
\mathcal{J}(q_{T},\tau)=\int_0^{\tau} t^2 dt
f(t)^2\left\langle\mu_5^2(t)\right\rangle |\mathcal{I}(q_{T}t)|^2\tilde{F}(q_T),\label{eq4.2.5}
\end{aligned}
\end{equation}
the integration is over proper time and
\begin{equation}
	\mathcal{I}(q_{T}t )=2\int_{0}^{\infty}e^{-iq_{T} t\cosh{\eta}}\cosh{\eta}\,d\eta = 2K_1(iq_{T} t),\label{eq4.2.7}
\end{equation}
with $K_1$ a modified Bessel function of the second kind. Finally,
\begin{eqnarray}
\tilde{F}(q_T) &=& \int d^2u
\mathcal{F}(u/\lambda) e^{i\bm{q_{T}\cdot \bm{u}}}
\nonumber\\
&=&2\pi\int_0^\infty u~du~\mathcal{F}(u/\lambda)
J_0(q_T u),
\label{eq:analeq2}
\end{eqnarray}
where we have used the assumption that $\mathcal{F}$ depends
only on $u$ while $\lambda$ is only a scale related to saturation momentum, as well as
\begin{equation}
\int_0^{2\pi}d\theta\, e^{i q_T u\cos\theta}
=2\pi J_0(q_T u).
\label{eq:analeq3}
\end{equation}

In order to extract the behavior of the photon spectrum at low $q_T$,
namely $q_T\ll g^2\mu$, we work at the limit $\bm q_T\rightarrow 0$ in Eq.~\eqref{eq:analeq2}. Hence, in this case
\begin{equation}
\tilde{F}(q_T)\approx 2\pi c\lambda^2,
\label{eq:analeq4}
\end{equation}
where $c$ is a dimensionless constant given by
\begin{eqnarray}
c &=&  \frac{1}{\lambda^2}
\int_0^\infty u~du~\mathcal{F}(u/\lambda),
\end{eqnarray}
and embeds the information about the shape of the 
$\mu_5$-correlator.
For $q_T\rightarrow 0$,  
$\tilde{F}(q_T)$ thus does not depend on $q_T$
and the only dependence of the spectrum on $q_T$
comes from $|\mathcal{I}(q_T t)|^2$, that in this limit reads
\begin{equation}
|\mathcal{I}(q_T t)|^2\approx \frac{4}{q_T^2 t^2},
\label{eq:analeq4.1}
\end{equation}
where we used~\cite{abra1965}
\begin{equation}
K_1(z)\approx \frac{1}{z},~~~|z|\rightarrow 0.
\label{eq:sfB2}
\end{equation}
Taking into account Eqs.~\eqref{eq:analeq4} 
and~\eqref{eq:sfB2}, we can write
\begin{equation}
|\mathcal{I}(q_T t)|^2\tilde{F}(q_T)
\approx
\frac{8\pi c \lambda^2}{q_T^2t^2},
\label{eq:analeq5}
\end{equation}
and
the spectrum~\eqref{eq:analeq1.1} in the
low-$q_T$ domain becomes
\begin{eqnarray}
\frac{dN_{\gamma}}{2\pi q_{T}dq_{T}dy}&=&
\mathcal{C}\mathcal{O} \frac{8\pi c \lambda^2}{q_T^2},
\label{eq:analeq6}
\end{eqnarray}
where
\begin{equation}
\mathcal{C}=
\frac{1}{2}
\alpha_e\left(\frac{N_c{\rm Tr}(Q^2)}{4\pi^3}\right)^2
(eB_0)^2 A_{T} \lambda_\tau,
\end{equation}
and
\begin{eqnarray}
\mathcal{O}&=&
  \int_0^\tau f^2(t) \left\langle\mu_5^2(t)\right\rangle d t.
\label{eq:analeqO}
\end{eqnarray}
The overall coefficient $\mathcal{O}$ contains the information
of the $u$-dependence of the $\mu_5$-correlator,
which does not affect the low-$q_T$ behavior of the photon
spectrum, as well as (and most importantly) of the
time duration of the magnetic field, via $f^2(t)$.
Moreover, we notice that in the low-$q_T$ regime the spectrum
$\propto 1/q_T^{2}$ in agreement with the result
of the full numerical
calculation shown in Fig.~\ref{Fig:photo3}.
The result~\eqref{eq:analeq6} agrees with what we
would have obtained using a simple 
$\lambda^2\delta^{(2)}(\bm x_T - \bm y_T)$ as $\mathcal{F}$ in Eq.~\eqref{eq4.2.3}:
this is reasonable, 
since small-$q_T$ photons probe
the large-$u$ part of the $\mu_5$-correlator,
and in this domain
the correlations are very small, so for all the practical
purposes correlations in 
this domain are not very different from those of an 
uncorrelated system.

Next we turn to the large-$q_T$ limit. For the purpose of extracting the large-$q_T$ behavior of the
photon spectrum,
we expand $\mathcal{F}$ in Eq.~\eqref{eq4.2.3} as
\begin{equation}
\mathcal{F}(u/\lambda) = e^{-u/\lambda}
\sum_{\ell=0}^na_{\ell}\left(\frac{u}{\lambda}\right)^\ell,
\label{eq:analeq8}
\end{equation}
where the exponential is taken in order to ensure the
convergence of the integral, at the same time, it also can guarantee that correlator die quickly at large distance. A direct calculation for small values of $\ell$ shows that in the large-$q_T$ limit
the terms with $\ell=0$ and $\ell=1$ dominate
in Eq.~\eqref{eq:analeq2}, as they
lead to $\tilde{F}\propto 1/q_T^3$, while the terms with
higher $\ell$ lead to $\tilde{F}\propto 1/q_T^\alpha$ with $\alpha\ge 5$.
Therefore, for the purpose of our analysis we can write Eq.~\eqref{eq:analeq8} as
\begin{equation}
\mathcal{F}(u/\lambda) = e^{-u/\lambda}
\left(
a_0 + a_1 \frac{u}{\lambda} \right).
\label{eq:analeq9}
\end{equation}
Moreover, both the terms with $a_0$ and $a_1$ lead to
$\tilde{F}\propto 1/q_T^3$, therefore in order to extract the $q_T$-dependence of the photon spectrum for large $q_T$ we can limit ourselves to consider only the first term in Eq.~\eqref{eq:analeq9}. Without loss of generality, we set $a_0 = 1$, which allows us to enforce $\mathcal{F}$ is trivial when $u=0$, i.e. $\mathcal{F}(0)=1$. Thus we write
\begin{equation}
\mathcal{F}(u/\lambda) = e^{-u/\lambda},
\label{eq:analeq10}
\end{equation}
and this gives
\begin{equation}
\tilde{F}(q_T)=\frac{2\pi\lambda^2}{(1+q_T^2\lambda^2)^{3/2}}
\approx \frac{2\pi\lambda^2}{q_T^3}.
\label{eq:analeq11}
\end{equation}
For the behavior of $|\mathcal{I}(q_T t)|^2$ for large $q_T$
we need to distinguish between the cases of small and large
proper time. If $q_T t\ll 1$ we can still use Eq.~\eqref{eq:analeq4.1}, which 
can be combined with Eq.~\eqref{eq:analeq11}
to give
\begin{equation}
|\mathcal{I}(q_T t)|^2\tilde{F}(q_T)
\approx
\frac{8\pi\lambda^2}{q_T^3}\frac{1}{q_T^2t^2}.
\label{eq:analeq12}
\end{equation}
On the other hand, for $q_T t \gg 1$ we can use~\cite{abra1965}
\begin{equation}
K_1(z)\approx \sqrt{\frac{\pi}{2z}}e^{-z},~~~|z|\rightarrow 
\infty,
\label{eq:analeq13}
\end{equation}
which gives, in the limit $q_T t \rightarrow\infty$,
\begin{equation}
|\mathcal{I}(q_T t)|^2\approx \frac{4\pi}{q_T t},
\label{eq:analeq14}
\end{equation}
and consequently
\begin{equation}
|\mathcal{I}(q_T t)|^2\tilde{F}(q_T)
\approx
\frac{8\pi^2\lambda^2}{q_T^3}\frac{1}{q_T t }.
\label{eq:analeq15}
\end{equation}
As is noticed, in the large-$q_T$ limit, the
behavior of the photon spectrum depends on the proper time: for small proper time the spectrum $\sim 1/q_T^5$,
while for large proper time the spectum $\sim 1/q_T^4$.
Therefore, in order to translate this information into a prediction for the photon spectrum, we can split the integration over proper time in Eq.~\eqref{eq4.2.5} into two ranges,
namely $(0,\tau_0)$ and $(\tau_0,\tau)$, with 
$g^2\mu\tau_0=O(1)$: for $\tau\le\tau_0$ we use
Eq.~\eqref{eq:analeq12} while for $\tau\ge\tau_0$
we use Eq.~\eqref{eq:analeq15}. Even through, the usability of above analysis will be weakened in domains where $q_T\tau\sim 1$, it is worthwhile to sacrifice some rigorousness for building up the connection between CME photon behavior and a scale related to $g^2\mu$ (further to the saturation momentum). Then it is easy to notice that if
$\tau\le\tau_0$ we have only the contribution $\sim1/q_T^5$,
while if $\tau\ge\tau_0$ we also need to include
the contribution $\sim1/q_T^4$ that dominates over the 
one $\sim 1/q_T^5$ for large $q_T$.
Eventually, we can summarize our results 
for the large-$q_T$ limit in the two ranges
as following form  
\begin{eqnarray}
\frac{dN_{\gamma}}{2\pi q_{T}dq_{T}dy}&\approx&\mathcal{C}\mathcal{O}
\frac{8\pi  \lambda^2}{q_T^5},\label{eq:analeq16}
\end{eqnarray}
on the other hand,
\begin{eqnarray}
\frac{dN_{\gamma}}{2\pi q_{T}dq_{T}dy}&\approx&
\mathcal{C}
\tilde{\mathcal{O}} \frac{8\pi^2  \lambda^2}{q_T^4},
\label{eq:analeq17}
\end{eqnarray}
with
\begin{eqnarray}
\tilde{\mathcal{O}}&=&
  \int_{\tau_0}^\tau tf^2(t) \left\langle\mu_5^2(t)
\right\rangle d t.
\label{eq:analeqOt}
\end{eqnarray}

It is interesting to notice that the behavior of the
photon spectrum for large $q_T$
depends on the duration of the
magnetic field, while the details of the shape
of the $\mu_5$-correlator do not affect such behavior. 
In particular,
if $\tau_B$ is very small, or if the proper time
$\tau$ at which the photon spectrum is computed is small,
then our analysis predicts that
the spectrum asymptotically $\sim 1/q_T^5$.
This is confirmed by our full numerical
calculations, see the case $\tau_B=0.1$ fm/c
in Fig.~\ref{Fig:photo2} and 
the case $\tau=0.1$ fm/c
in Fig.~\ref{Fig:photo3}.
On the other hand, if the magnetic field survives for a 
long enough time, and if the spectrum
is computed at larger proper times,
then the photon production takes place also for
$\tau\ge \tau_0$, hence the asymptotic behavior of
the spectrum changes to $\sim 1/q_T^4$.
This is also in agreement with the results shown in 
Fig.~\ref{Fig:photo2} and in Fig.~\ref{Fig:photo3}.
We thus conclude that the asymptotic form of the spectrum
depends on the time at which photons are produced,
which in last analysis depends on the duration of the
magnetic field in the collisions.

\subsection{The elliptic flow}\label{secv2}
A convenient way of characterizing the various patterns of anisotropic flow is to use a Fourier expansion of the differential distribution:
\begin{equation}
	q_0\frac{dN_{\gamma}}{dq^3}=\frac{1}{2\pi}\frac{dN_{\gamma}}{q_{T}dq_{T}dy}\biggl(1+2\sum_{n=1}^{\infty}v_n\cos[n(\phi-\psi_{RP})]\biggr),\label{eq4.4.1}
\end{equation}
where $\phi$ is the azimuthal angle, and $\psi_{RP}$ is the reaction plane angle. For the purpose of this section, it is enough to assume that $\psi_{RP} = 0$. 

It is interesting to notice that the photon distribution~\eqref{eq4.2.1} is anisotropic in momentum space, and is characterized by a finite elliptic flow, $v_2$. As a matter of fact, at midrapidity it depends on the overall factor
\begin{equation}
1-\frac{q_y^2}{\bm{q}^2}=\cos^2{\phi}=\frac{1+\cos{2\phi}}{2},\label{eq4.4.2}
\end{equation}
which, according to Eq.~\eqref{eq4.4.1},
and taking into account that
$\langle\zeta(\bm q_T)\rangle$ 
in Eq.~\eqref{eq4.2.1} depends on $|\bm q_T|$ only,
immediately gives
\begin{equation}
v_2=\frac{1}{2}.
\label{eq:analeq18}
\end{equation}

This large $v_2$ is produced instantaneously, and is not related to an actual anisotropic flow of the medium. This is different from the $v_2$ of the QGP, that arises from the anisotropic flow of the bulk and needs time to be formed~\cite{Ruggieri:2013bda,Ruggieri:2013ova}.
The net number of CME photons is relatively smaller than the ones produced by other sources during the entire evolution of the medium, see Fig.~\ref{Fig:photo4}. 
\begin{figure}[t!]
	\centering
	\includegraphics[width=\linewidth]{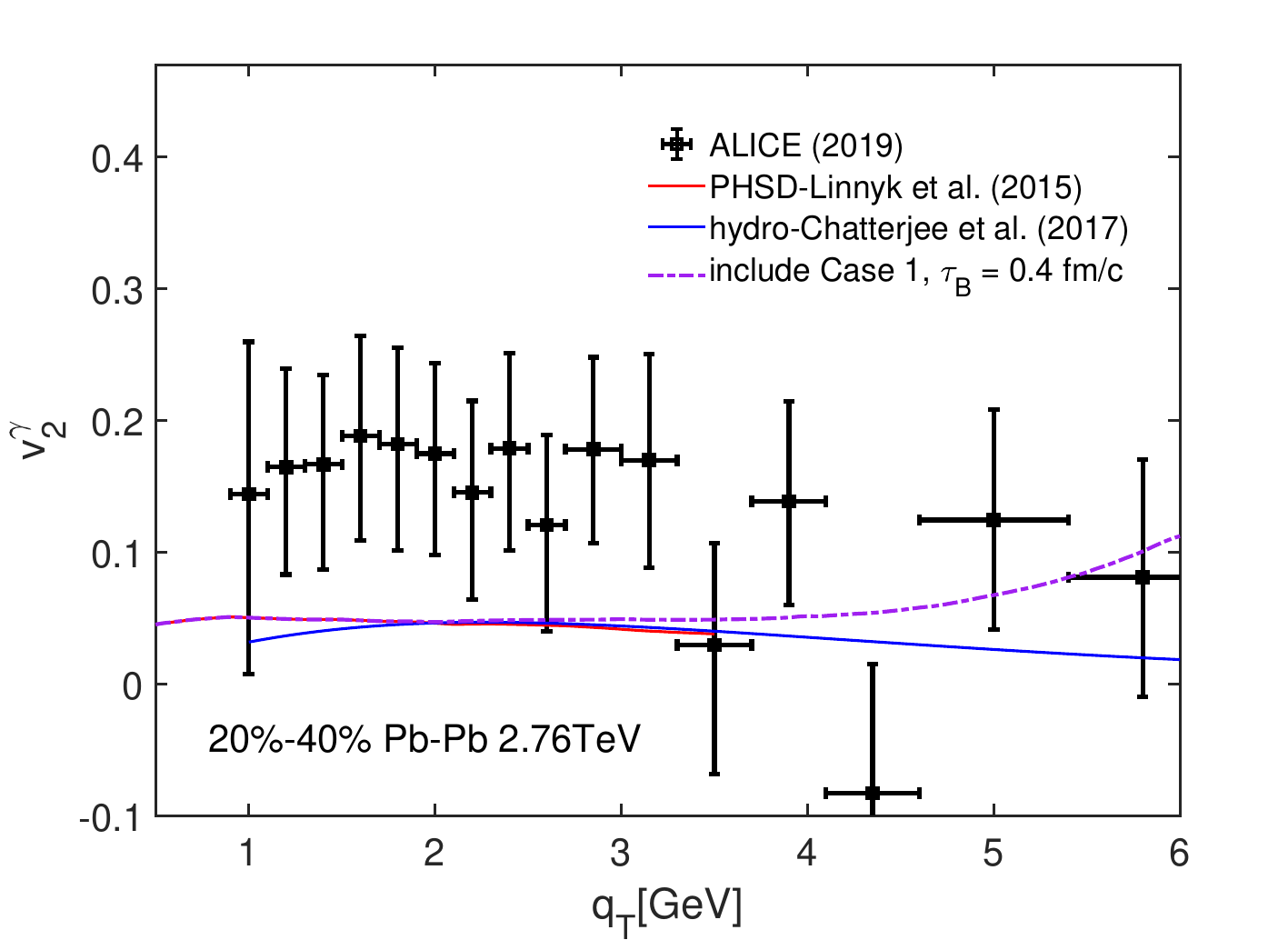}
	\caption{Comparison of the elliptic flow between ALICE measurement \cite{ALICE:2018dti}, PHSD model \cite{Linnyk:2015tha}, hydrodynamic approach \cite{Chatterjee:2017akg} and the weighted result by including CME photons. }
	\label{Fig:photo5}
\end{figure}
However, given the large value of $v_2$ carried by these photons, it is likely that they will give some contribution to the total $v_2$. In Fig~\ref{Fig:photo5}, we compare the ALICE measurement  \cite{ALICE:2018dti} with PHSD computation \cite{Linnyk:2015tha} (solid red line) and hydrodynamic approach \cite{Chatterjee:2017akg} (solid blue line). To visualize the contribution of CME photon (dot-dashed purple line), we weight our result on top of the two current models, the weight is defined as
\begin{equation}
	v^{\gamma}_{2} = \frac{\sum_{i} \frac{dN_{\gamma}^{i}}{2\pi q_Tdq_Td\eta}v_{2}^{i}}{\sum_{i}\frac{dN_{\gamma}^{i}}{2\pi q_Tdq_Td\eta}},\qquad i\in \{\rm{PHSD, hydro, CME}\}, \label{eq:weightv2}
\end{equation}
where we use index $i$ to label the source of photon. It is inspiring to find that the contribution from CME photon forwards the estimation of the two models to measurements in larger $q_T$ region. 

\section{conclusions and outlook}\label{sec5}

We carried out an event-by-event study of direct photons produced by the Chiral Magnetic Effect (CME) in the early stage of high-energy nuclear collisions, modeled by an evolving glasma. Assuming a strong magnetic background, the photons within our model are produced by an inverse Primakoff effect in which the chiral chemical
potential, $\mu_5$,  is related to a space-time dependent $\theta$ angle and is converted to on-shell photons via the QED anomaly. The chiral chemical potential, $\mu_5$, is dynamically produced within the present model by the QCD anomaly, as described by Eqs.~\eqref{eq2.3.1} and~\eqref{eq3.11}. Although the ensemble average of $\mu_5$ locally vanishes, its fluctuations are nonzero and can be related to the photon production rate. We assumed a uniform, time-dependent magnetic field decaying on a time scale $\tau_B$, that we took to be of the order of a fraction of fm/c and treated as a free parameter. Furthermore, we analyzed two different forms for the time dependence of the magnetic field. Both the time dependence and the values of $\tau_B$ were taken from previous studies~\cite{Sun:2020wkg} in which the role of the electromagnetic field in heavy ion collisions was studied in relation
to the directed flow.

We found that $\sqrt{\langle\mu_5^2\rangle}$ can be quite large in the early stage, reaching the order of the saturation scale, $\sqrt{\langle\mu_5^2\rangle}=O(\mathrm{GeV})$. We then computed the spectrum of CME photons and investigated how this spectrum depends on the value of $\tau_B$ as well as on the time dependence of the magnetic field. Our findings showed that increasing $\tau_B$ leads to an enhancement in CME photon production. Furthermore, these photons are produced quickly, predominantly within the very early stage. In fact, we found that the CME photon production almost stops after about $0.5$ fm/c. Despite minor quantitative differences, we observed that the specific choice of the magnetic field's decay profile, while influential on other observables as reported in Ref.~\cite{Sun:2020wkg}, does not significantly impact the production of photons via the CME.

Moreover, we extracted the analytical behavior of the CME photons for small and large transverse momentum, $q_T$. Interestingly, we established a connection between the large-$q_T$ behavior of the photon spectrum and the duration of the magnetic field in the collisions. Notably, we found that this behavior remains independent of the specific form of the $\mu_5$-correlator: while the latter might influence the overall coefficient of the asymptotic expansion of the spectrum for large $q_T$, it does not alter the inverse power of $q_T$ characterizing the spectrum in the large-$q_T$ domain. We then compared the CME photon rate with that of other sources of photons in the early stage, including prompt photons and those produced within hydrodynamic model with NLO pQCD prediction. Although the net photon yield from the CME mechanism is relatively smaller than that from other mechanisms, we observed that CME photons exhibit a substantially large elliptic flow, specifically $v_2=1/2$. Consequently, they could contribute significantly to the total elliptic flow of photons generated in relativistic heavy-ion collisions. To visualize this contribution, we weighted result of CME photon on top of PHSD predictions and hydrodynamic results with respect to Eq.~\eqref{eq:weightv2}. As was found, when compared to the predictions by PHSD model and hydrodynamic model, contributions from CME photon on total $v_2$ could be significant especially in a relatively large $q_T$ domain. 

\begin{acknowledgments}
The author would like to acknowledge Kenji Fukushima, Marco Ruggieri, Xu-Guang Huang and Defu Hou for inspirations and insightful comments. And special thank to Marco Ruggieri for instructive discussions and suggestions during the analysis of numerical results. At same time, the author would also like to thank Lucia Oliva for fruitful discussions during this work.
\end{acknowledgments}

\bibliography{BIBL}

\end{document}